# A Real-World Demonstration of Machine Learning Generalizability: Intracranial Hemorrhage Detection on Head CT

Hojjat Salehinejad[1,2], Jumpei Kitamura[3], Noah Ditkofsky[4,5], Amy Lin[4,5], Aditya Bharatha[1,4,5], Suradech Suthiphosuwan[4,5], Hui-Ming Lin[1], Jefferson R. Wilson[1,5,6], Muhammad Mamdani[1,5,7-8], and Errol Colak[1,4,5,*]

1 Li Ka Shing Centre for Healthcare Analytics Research & Training, St. Michael's Hospital, Toronto, Canada
2 Department of Electrical & Computer Engineering, University of Toronto, Toronto, Canada
3 Fujisawa-city, Kanagawa, Japan
4 Department of Medical Imaging, St. Michael's Hospital, Unity Health Toronto, Toronto, Canada
5 Faculty of Medicine, University of Toronto, Toronto, Canada
6 Division of Neurosurgery, Department of Surgery, University of Toronto, Toronto, Canada
7 Leslie Dan Faculty of Pharmacy, University of Toronto, Toronto, Canada
8 Dalla Lana Faculty of Public Health, University of Toronto, Toronto, Canada

* Correspondence to: Dr. Errol Colak, Department of Medical Imaging, St Michael's Hospital, Unity Health Toronto, 30 Bond Street, Toronto, ON, Canada M5B 1W8. E-mail: errol.colak@unityhealth.to

## ABSTRACT

Machine learning (ML) holds great promise in transforming healthcare. While published studies have shown the utility of ML models in interpreting medical imaging examinations, these are often evaluated under laboratory settings. The importance of real world evaluation is best illustrated by case studies that have documented successes and failures in the translation of these models into clinical environments. A key prerequisite for the clinical adoption of these technologies is demonstrating generalizable ML model performance under real world circumstances.

The purpose of this study was to demonstrate that ML model generalizability is achievable in medical imaging with the detection of intracranial hemorrhage (ICH) on non-contrast computed tomography (CT) scans serving as the use case. An ML model was trained using 21,784 scans from the RSNA Intracranial Hemorrhage CT dataset while generalizability was evaluated using an external validation dataset obtained from our busy trauma and neurosurgical center. This real world external validation dataset consisted of every unenhanced head CT scan (n = 5,965) performed in our emergency department in 2019 without exclusion. The model demonstrated an AUC of 98.4%, sensitivity of 98.8%, and specificity of 98.0%, on the test dataset. On external validation, the model demonstrated an AUC of 95.4%, sensitivity of 91.3%, and specificity of 94.1%. Evaluating the ML model using a real world external validation dataset that is temporally and geographically distinct from the training dataset indicates that ML generalizability is achievable in medical imaging applications.

# Introduction

Intracranial hemorrhage (ICH) is a source of significant morbidity and mortality [1,2]. It is a frequently encountered clinical problem with an overall incidence of 24.6 per 100,000 person-years [3]. A non-contrast computed tomography (CT) scan of the head is the most common method used to diagnose ICH as it is fast, accurate, and widely available. Since nearly half of ICH related mortality occurs within the first 24 hours [4], rapid and accurate diagnosis is critical if interventions that can improve patient outcomes are to be successful [5-8].

In high volume clinical radiology settings with complex patients and frequent interruptions, significant delays between patient imaging and imaging interpretation are often unavoidable. Inevitably, this delay will impact the time required to identify patients with critical or life-threatening findings [9]. Machine learning (ML) models have been proposed as an approach to automatically triage and prioritize medical imaging studies [10]. Multiple investigators have demonstrated the accuracy of ML models in detecting ICH on non-contrast CT scans [11-15]. However, many previously published investigations have not evaluated performance of these ML models in real world, high volume clinical environments. The importance of real world evaluation is best demonstrated by case studies which have shown failures in translation from laboratory to clinical settings due to a variety of sociotechnical factors [16]. Another limitation of many of these studies is a common source of the training, validation, and test datasets.

A demonstration of generalizable ML model performance on real world data is necessary prior to the adoption of these tools. In this paper, we developed an ML model for ICH detection in non-contrast CTs of the head and examined generalization performance in the real world setting of a major neurosurgical and trauma center. To our knowledge, this is the first study to both develop and assess generalization performance of an ML model for ICH detection.

# METHODS

The following methods were carried out in accordance with relevant institutional guidelines and regulations. The study was approval by the St. Michael's Hospital Research Ethics Review Board with a waiver of informed consent.

**Training Dataset**

The Radiological Society of North America (RSNA) Intracranial Hemorrhage CT dataset [17] was used for ML model training. This multi-institutional and multi-national dataset is composed of head CTs and annotations of the five types of intracranial hemorrhage. Each CT image in this dataset was annotated by a neuroradiologist for the presence or absence of epidural (EDH), subdural (SDH), subarachnoid (SAH), intraventricular (IVH), and intraparenchymal (IPH) hemorrhage. This dataset consists of 874,035 images with class imbalance amongst the types of ICH (Table 1).

Table 1. Distribution of examination labels in the training, test, and external validation datasets according to hemorrhage types. The number of labels exceeds the actual number of examinations as more than one label may have been applied to each CT scan.

|  | Training | Test | External |
|---|---|---|---|
| Any hemorrhage type | 8,889 | 1,243 | 674 |
| Epidural | 354 | 23 | 25 |
| Subdural | 3,814 | 503 | 367 |
| Subarachnoid | 3,936 | 528 | 288 |
| Intraventricular | 3,692 | 616 | 128 |
| Intraparenchymal | 5,324 | 758 | 287 |
| No hemorrhage | 21,784 | 3,528 | 674 |

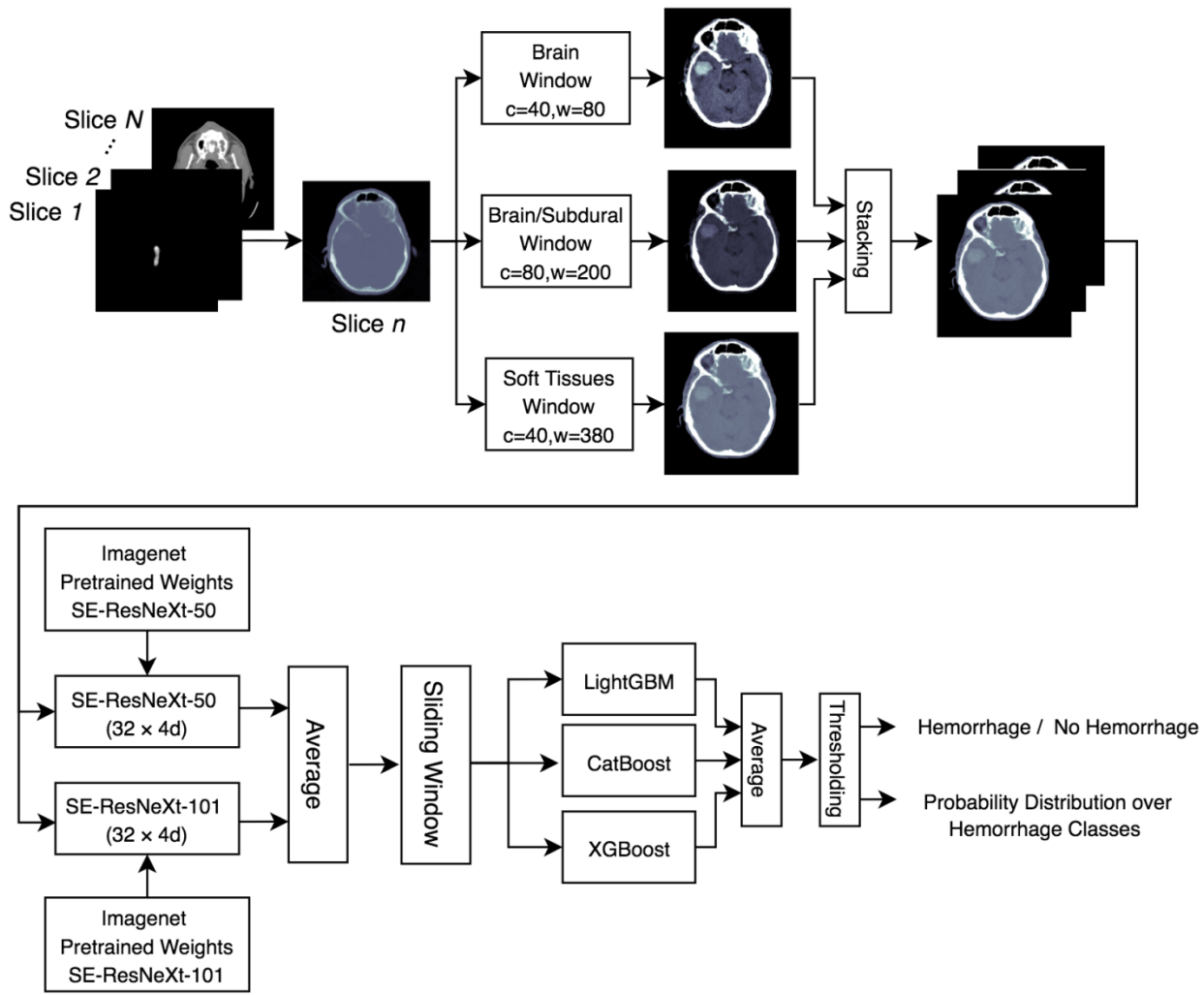

Figure 1. Architecture of the ML model.

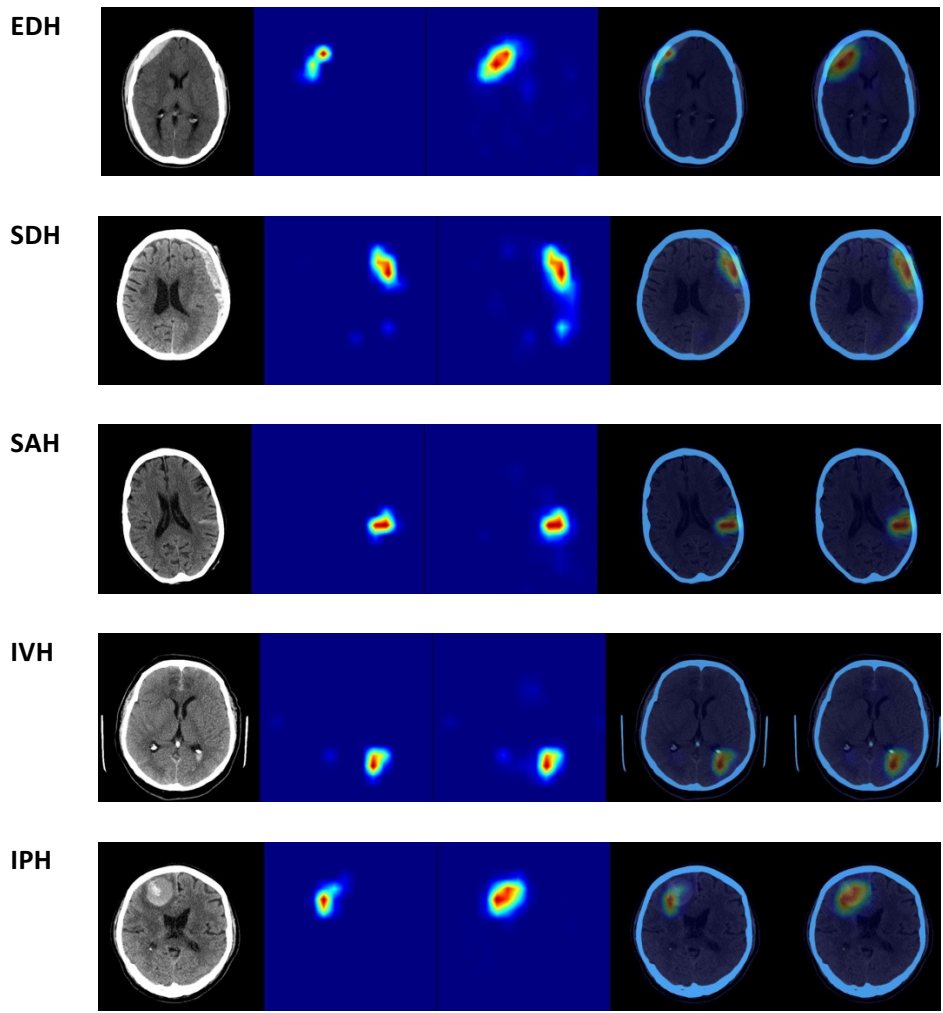

**Figure 2.** Visualization of feature maps from layer 4 (the layer before adaptive averaging pooling) in SE-ResNeXt-50 (32x4d). Left: Input head CT image; Middle left: GradCAM heat map; Middle: GradCAM++ heat map; Middle right: GradCAM result superimposed on CT image; Right: GradCAM++ result superimposed on CT image.

**Model Development**

An overview of the ML model is presented in Figure 1. The main steps are as follows:

1. Adjustment of the window center and width of each CT image;
2. Feature extraction from each image;
3. Incorporation of spatial dependencies between images along the craniocaudally axis;
4. Thresholding inference results to generate a binary decision and a probability distribution over the 5 types of ICH.

A CT scan of the head is represented as $S = (S_1,...,S_N)$ where $N$ is the total number of images in the scan. Each image $S_n$ is passed through three window center and width adjustment filters to enhance differences between blood, brain parenchyma, cerebrospinal fluid, soft tissues, and bone [18] as presented in Figure 1. The three enhanced images are then stacked and passed to two deep convolutional neural networks (DCNN) with three input channels which are SE-ResNeXt-50 and SE-ResNeXt-101, pre-trained on ImageNet [19]. Each DCNN model produces a probability distribution over the target data classes for each $S_n$ and their average is defined as the vector $p_n = (p_n^{(1)}, p_n^{(2)}, p_n^{(3)}, p_n^{(4)}, p_n^{(5)})$, where indexes 1 to 5 refer to the EDH, SDH, SAH, IVH, and IPH classes, respectively. An ensemble of the probability distributions generated by the DCNNs was used to reduce the variance of predictions. In order to incorporate spatial dependency between axial images, a sliding window module takes the probability vectors of $\Delta S$ images from each side of image $S_n$ as $P_n = (p_{n-\Delta S}, p_{n-\Delta S+1}, ..., p_n, p_{n+1}, ... p_{n+\Delta S})$. The prediction $P_n$ is then enhanced by incorporating inter-slice dependencies using an ensemble of the LightGBM, CatBoost, and XGBoost gradient boosting models [20]. The average of the ensemble model produces a probability distribution over the 5 hemorrhage types for the slice $S_n$. This distribution is passed to a set of thresholds where if at least the predicted probability of one hemorrhage type is more than or equal to its corresponding threshold, the output label will be positive for ICH.

A Bayesian optimizer [21] was used to determine the probability thresholds ($T_{EDH}$ = 0.47, $T_{SDH}$ = 0.37, $T_{SAH}$ = 0.45, $T_{IVH}$ = 0.37, and $T_{IPH}$ = 0.20) that maximize the AUC score when generating binary (positive/negative) decisions. Visualization of predicted areas of ICH was performed using feature maps from layer 4, the layer before adaptive averaging pooling, in the SE-ResNeXt-50 (32x4d) model using GradCAM and GradCAM++ methods [22] (Figure 2). This visualization is used to confirm that the ML model is capable of detecting areas of hemorrhage without performing any geometrical prepossessing (e.g. image registration, noise removal) on the input head CT images even in the presence of suboptimal patient positioning or other artifacts.

**Model Training and Evaluation**

The training portion of the RSNA Intracranial Hemorrhage CT dataset of 752,803 images (21,784 examinations) was used to train the DCNNs and divided into 8 stratified folds. Images from the same patient were grouped into the same fold by using the patient identifier embedded in DICOM metadata. This prevents a potential information leak during cross-validation as neighboring images within a CT scan may resemble each other and are more likely to share the same class labels. Each DCNN model was trained and cross-validated on these 8 folds. The training hyper-parameters of the DCNNs were set to a mini-batch size of 32, training epoch of 4, and adaptive learning rate with initial rate of $1\times10^{-4}$ with an Adam optimizer [23]. The checkpoints from the 3rd and 4th epochs were used to make out-of-fold predictions and were then averaged. These out-of-fold predictions were used as meta features for training gradient boosting models. Cross-validation was performed on the same 8 folds.

The model was evaluated on the 3,528 examinations that compose the test set of the RSNA Intracranial Hemorrhage CT dataset. Log loss performance during training and validation was determined for SE-ResNeXt50-32x4d and SE-ResNeXt101-32x4d for each fold and epoch. In addition, log loss was determined for LightGBM, Catboost, and XGB, as well as their average as an ensemble, for each hemorrhage type. A confusion matrix was constructed by comparing the ground truth of each CT scan to the ML model prediction.

**Evaluation of Model Generalizability**

The demonstration of generalizability of model performance requires external validation using data which is ideally both temporarily and geographically distinct from that used to train a model [24]. As a busy neurosurgical and trauma center in one of the world's most diverse cities, the data from our institution is well suited for the purposes of external validation. In order to capture a real-world distribution of patients, we included every unenhanced head CT performed on emergency department patients over the course of 1 year without any exclusion criteria.

*External Validation Dataset*

The hospital's radiology information system (syngo, Siemens Medical Solutions USA, Inc., Malvern, PA) was searched using Nuance mPower (Nuance Communications, Burlington, MA) for emergency patients that underwent a non-contrast CT scan of the head between January 1 and December 31, 2019. Every CT which included non-contrast imaging of the head acquired at 2.5 or 5.0 mm slice thickness was included in this study. All examinations were performed on a 64 row multi-detector CT scanner (Revolution, LightSpeed 64, or Optima 64, General Electric Medical Systems, Milwaukee, WI).

The ground truth was established by having each CT scan labeled as positive or negative for ICH by a trained research assistant who reviewed the associated radiology report. A total of 5,965 (674 positive, 5,291 negative) head CT examinations from 5,536 patients (2,600 female, 3,365 male; age range 13-101 years; mean age 58.2 ± 20.4 years) were included in this study.

Scans that were positive for ICH were further classified for the presence of the 5 types of ICH using the same radiology report. A random sample of 600 reports and CT scans (64 positive and 536 negative) were reviewed by a radiologist to validate the report labeling process. All positive and negative scans were correctly classified by the research assistant at the patient level. For the positive scans, 314 of 320 (98.1%) labels detailing the types of ICH were correctly labeled. A total of 103 of 105 ICH subtype positive labels were correct, 2 were reclassified, and 5 were added after radiologist review.

*Evaluation*

ML model predictions were compared to the ground truth for each scan at the patient level and for each type of ICH. A panel of three neuroradiologists reviewed each CT scan where the ground truth label based on the clinical radiology report was discrepant with the ML model prediction. This review allowed us to identify cases where a radiologist missed ICH that was correctly detected by the ML model and cases of "over-calling" by a radiologist. A majority vote served as consensus for the review of these cases. CT scans that were deemed equivocal for ICH by the panel despite the availability of prior and follow-up imaging were treated as positive cases in evaluating ML model performance. The rationale for this decision is that equivocal cases should be flagged by a triaging system for urgent review by a radiologist.

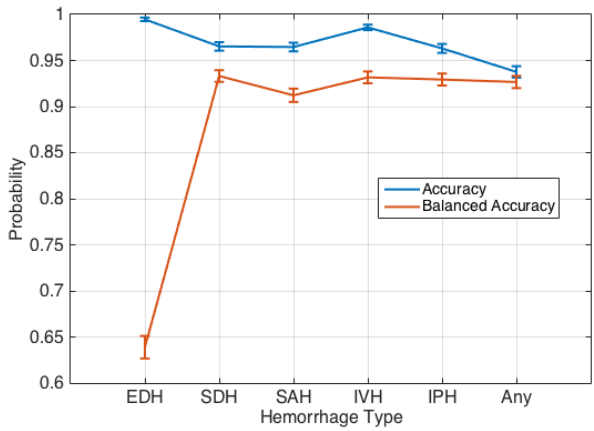 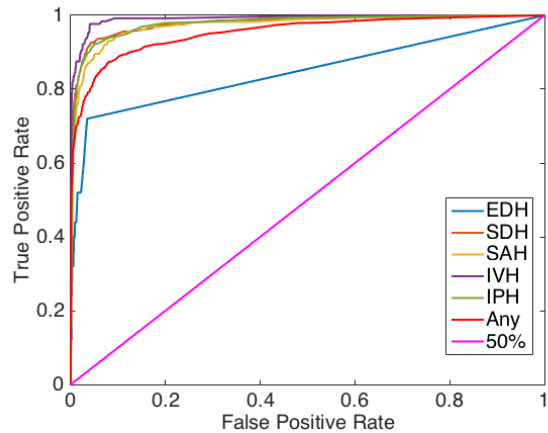

**Figure 3. Probability estimates and 95% confidence intervals of hemorrhage types with respect to the accuracy and balanced accuracy measures.**

**Figure 4. Receiver operator curves (ROC) of hemorrhage type from the external validation dataset.**

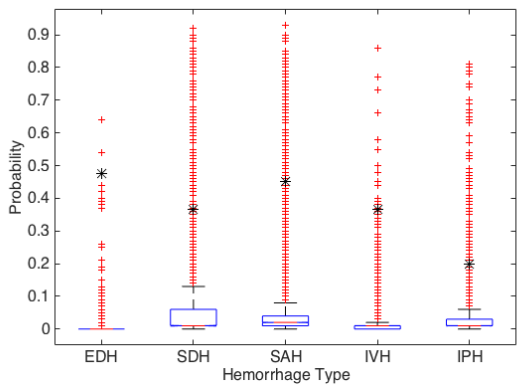 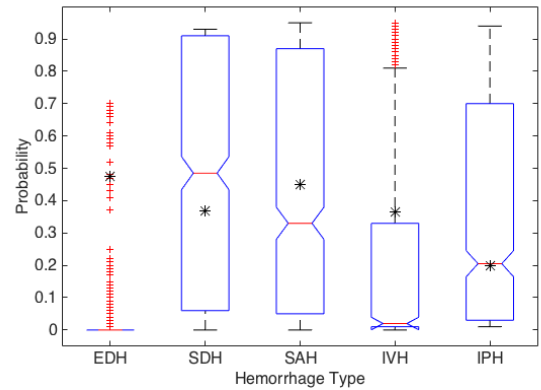

**Figure 5. Probability distribution of the predicted labels for ground truth negative and positive cases. The central red line indicates the median, the bottom and top edges of the box indicate the 25th and 75th percentiles, respectively, and the whiskers extend to the most extreme data points not considered outliers. The outliers are plotted individually using the red "+" symbol and the found threshold by Bayesian optimizer is plotted using the black "*" symbol. Cases with a probability higher than the threshold are counted toward the corresponding positive and negative label.**

From a probability theory perspective, we can model each CT scan as an independent event with respect to a hemorrhage type, that is either is positive (success) or negative (failure). For a one-year sample of data, this set of events can be modeled as a Bernoulli process [25]. A binomial distribution for a large number of samples can be approximated by a Gaussian distribution using the Central Limit Theorem [25,26] and be confidently used to calculate the confidence intervals (CI).

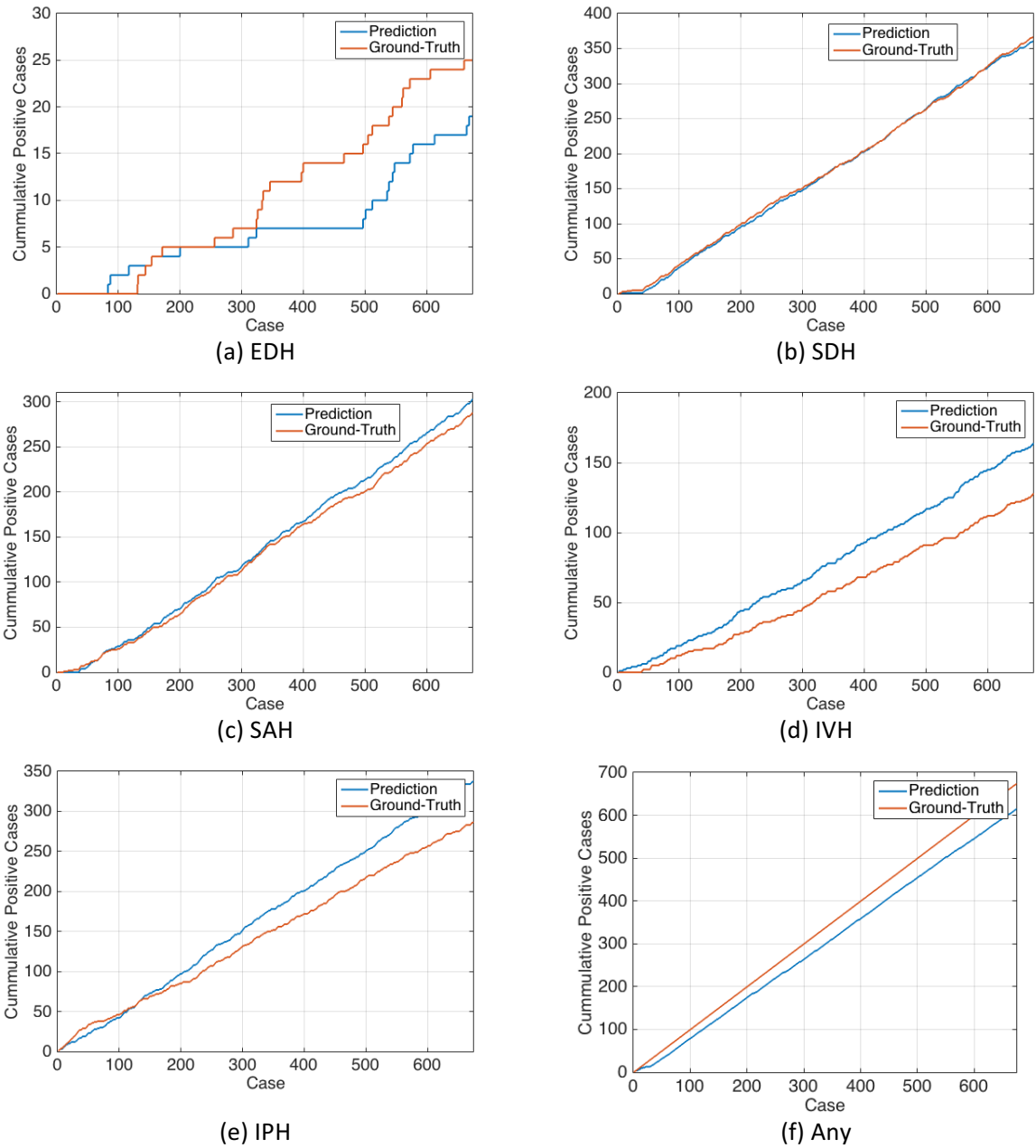

**Figure 6. Cumulative value of positive hemorrhage cases.**

In order to visually illustrate the performance of the ML model compared to the ground-truth per scan at different scan intervals, cumulative positive case versus ground truth plots were generated at the patient level and for each type of ICH. If a CT scan is positive for a hemorrhage type, one is added to the cumulative value and if it is negative, zero is added. More divergence of the curves means less agreement between the ML model and the ground-truth. The difference at the last index is the number of scans where the ML model has made errors. If overall, the prediction curve is above the ground-truth curve, it means the ML model has over-called and if the prediction curve is below the ground-truth curve, the ML model has failed to diagnose cases with that specific type of hemorrhage.

Table 2. ML performance in detecting ICH on the test set. TP: True positive; FN: False negative; TN: True negative; FP: False positive; SEN: Sensitivity; SPEC: Specificity; PPV: Positive predictive value; NPV: Negative predictive value; AUC: Area under the receiver operating curve; Acc: Accuracy; BAcc: Balanced accuracy; MCC: Matthews correlation coefficient; F1: F1 score. All the values except TP, FN, TN, and FP are in percent.

| Hemorrhage | TP | FN | TN | FP | SEN | SPEC | PPV | NPV | AUC | Acc | BAcc | MCC | F1 |
|---|---|---|---|---|---|---|---|---|---|---|---|---|---|
| EDH | 5 | 18 | 3493 | 2 | 21.5 | 99.9 | 71.4 | 99.5 | 60.8 | 99.4 | 60.8 | 39.2 | 33.3 |
| SDH | 424 | 79 | 2969 | 46 | 84.3 | 98.5 | 90.2 | 97.4 | 91.4 | 96.5 | 91.4 | 85.2 | 87.2 |
| SAH | 406 | 122 | 2952 | 38 | 76.9 | 98.7 | 91.4 | 96.0 | 87.8 | 95.5 | 87.8 | 81.3 | 83.5 |
| IVH | 574 | 42 | 2869 | 33 | 93.2 | 98.9 | 94.6 | 98.6 | 96.0 | 97.9 | 96.0 | 92.6 | 93.9 |
| IPH | 713 | 45 | 2713 | 47 | 94.1 | 98.3 | 93.8 | 98.4 | 96.2 | 97.4 | 96.2 | 92.3 | 93.9 |
| Any | 1228 | 15 | 2230 | 45 | 98.8 | 98.0 | 96.5 | 99.3 | 98.4 | 98.3 | 98.4 | 96.3 | 97.6 |

Table 3. ML performance in detecting ICH on the external validation set. TP: True positive; FN: False negative; TN: True negative; FP: False positive; SEN: Sensitivity; SPEC: Specificity; PPV: Positive predictive value; NPV: Negative predictive value; AUC: Area under the receiver operating curve; Acc: Accuracy; BAcc: Balanced accuracy; MCC: Matthews correlation coefficient; F1: F1 score. All the values except TP, FN, TN, and FP are in percent.

| Hemorrhage | TP | FN | TN | FP | SEN | SPEC | PPV | NPV | AUC | Acc | BAcc | MCC | F1 |
|---|---|---|---|---|---|---|---|---|---|---|---|---|---|
| EDH | 7 | 18 | 5926 | 14 | 28.0 | 99.8 | 33.3 | 99.7 | 84.7 | 99.5 | 63.9 | 30.3 | 30.4 |
| SDH | 329 | 38 | 5429 | 169 | 89.7 | 97.0 | 66.1 | 99.3 | 98.0 | 96.5 | 93.3 | 75.3 | 76.1 |
| SAH | 246 | 42 | 5508 | 169 | 85.4 | 97.0 | 59.3 | 99.2 | 97.4 | 96.5 | 91.2 | 69.5 | 70.0 |
| IVH | 112 | 16 | 5769 | 68 | 87.5 | 98.8 | 62.2 | 99.7 | 99.2 | 98.6 | 93.2 | 73.1 | 72.7 |
| IPH | 256 | 31 | 5489 | 189 | 89.2 | 96.7 | 57.5 | 99.4 | 97.9 | 96.3 | 92.9 | 69.9 | 70.0 |
| Any | 615 | 59 | 4978 | 313 | 91.3 | 94.1 | 66.3 | 98.8 | 95.4 | 93.8 | 92.7 | 74.5 | 76.8 |

## RESULTS

Evaluation of ML model performance on the test dataset revealed an AUC of 98.4%, a balanced accuracy of 98.4%, an imbalanced accuracy of 98.3%, sensitivity of 98.8%, specificity of 98.0%, positive predictive value of 96.5% and negative predictive value of 99.3% for ICH detection (Table 2).

ML model performance was then evaluated on the external validation dataset which revealed an AUC of 95.4%, a balanced accuracy of 92.7%, an imbalanced accuracy of 93.8%, sensitivity of 91.3%, specificity of 94.1%, positive predictive value of 66.3% and negative predictive value of 98.8% for ICH detection (Table 3). The 95% CI with respect to the balanced accuracy score for each hemorrhage class is EDH (± 1.22%), SDH (± 0.63%), SAH (± 0.72%), IVH (± 0.64%), IPH (± 0.65%), and ICH (± 0.66%) (Figure 3). The 95% CIs of EDH show a 35.58% difference between the CI of the accuracy and balanced accuracy scores. This indicates high generalization performance of the ML model for all types of ICH except EDH. A receiver operating characteristic (ROC) curve was generated using the external set (Figure 4).

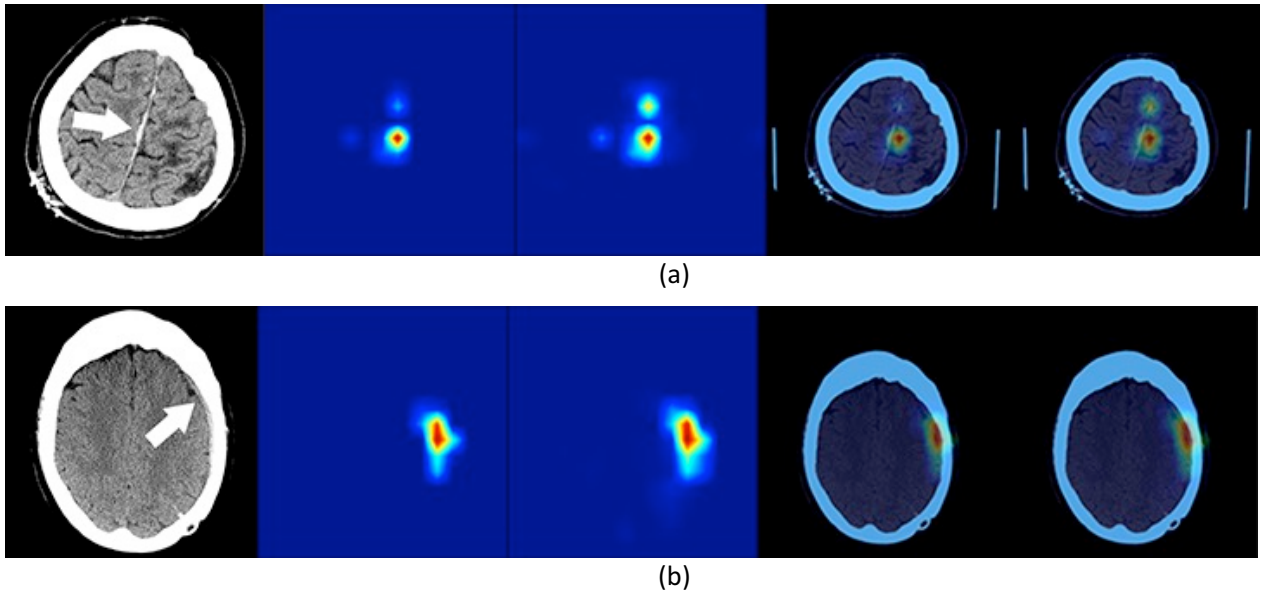

**Figure 7. (a) and (b):** Two examples of SDH that were missed by a radiologist but detected by the ML model. Visualization of feature maps from layer 4 (the layer before adaptive averaging pooling) in SE-ResNeXt-50 (32x4d). Left: Input head CT image; Middle left: GradCAM heat map; Middle: GradCAM++ heat map; Middle right: GradCAM result superimposed on CT image; Right: GradCAM++ result superimposed on CT image. The arrows indicate the SDH.

Figure 5 shows the distribution of predicted hemorrhage probability by the ML model for the external validation dataset at the patient level. This figure shows that the probability distribution of prediction for both negative and positive EDH cases is very similar. Figure 6 shows the cumulative positive cases between the ML model prediction and ground-truth. The ML model has under-called (i.e. missed) cases of EDH while over-calling SAH, IVH, and IPH. For SDH, the two curves are more aligned than the other hemorrhage types and represents the highest agreement between the ML model and ground-truth. The accuracy results in Table 3 express a similar conclusion.

A panel of neuroradiologists reviewed the CT scans of patients which were classified as false negative and false positive. Following this review, 17 of the 59 false negative and 16 of the 313 false positive predictions by the ML model were considered equivocal despite the availability of prior and follow-up imaging. Two cases of ICH were correctly detected by the ML model but missed when reported by a radiologist (Figure 7).

## DISCUSSION

In this study, we have shown that an ML model is able to demonstrate high generalizable performance in the detection of ICH. While many studies on ICH detection report high accuracy, a deeper examination shows that many of these studies suffer from limitations that may impede translation of ML models into real world clinical environments. For example, data from a common institution is often used for ML model training, validation, and testing. Many prior studies evaluate model performance on curated datasets that may not reflect the prevalence and variety of ICH encountered in clinical practice. Furthermore, investigators often do not specify the method used to curate such datasets.

When ML models are tested in real world environments, the sample size and evaluation period is often limited while the inclusion and exclusion criteria may not be clearly defined. After initial studies showing great performance, follow-up studies have shown that some ML models display lower accuracy and higher false positive rates in different clinical environments [27,28]. The RSNA and the American College of Radiology have recently expressed concern that many commercially available ML algorithms have failed to demonstrate comprehensive generalizability in heterogeneous patient populations, radiologic equipment, and imaging protocols [29].

We believe that we help address some of these concerns by demonstrating high model performance in a large heterogeneous dataset of head CTs performed over the course of one year in a busy neurosurgical and trauma centre in one of the world's most diverse cities [30]. This dataset did not exclude any emergency department patients irrespective of image quality and the presence of artifacts (e.g. motion, streak, etc.). Furthermore, the data used to train the ML model was distinct from our institutional dataset which shows that ML model generalizability can be achieved. The level of accuracy demonstrated by the ML model supports its use as a triage system, a second reader, or as part of a quality assurance system.

We considered equivocal cases for ICH as positive as we believe these cases should be flagged for urgent radiologist review. This decision had the impact of decreasing the reported performance of the ML model and an increase in the number of false negative classifications. In terms of false negative cases, a substantial number were thin subdural hematomas. The clinical significance of not detecting these hematomas is not certain but prior studies have suggested that a large proportion of small extra-axial collections do not require intervention [31]. The false positive rate we encountered translates into less than one case per day which is a trivial increase in radiologist workload and would not have a significant impact in delaying the review of other imaging studies. In fact, many of these false positive cases included mimickers of ICH such as brain neoplasms and diffuse hypoxic ischemic injury that represent significant pathology.

The ML model detected ICH on two scans which were missed by the interpreting radiologist in our subspecialty academic radiology practice environment. With "real-world" error rates in the interpretation of head CTs ranging between 0.8 and 2.0% [32,33], ML tools may have an important role to play in quality assurance or as a second reader. This could be particularly important in clinical environments with limited access to neuroradiology expertise.

We have shown that the probability distribution of prediction for both negative and positive EDH cases is very similar. This can be justified by the threshold found by the Bayesian optimizer where the threshold is very close to 0.5 (i.e. that is 50% chance of being EDH). This observation shows the bias of the ML model toward other hemorrhage types due to the limited number of training samples, which is a common problem in training ML models on medical images [34,35]. Potential solutions in addressing limited training data include image augmentation through geometrical transformations [35] and image synthesis [34,36].

This study has several limitations. The model was trained on the RSNA Intracranial Hemorrhage CT dataset. Images with EDH represent only a tiny fraction of this dataset which is reflected in the poorer performance of our model in detecting EDH. This issue could be mitigated by augmenting the amount of EDH training data through computer mediated techniques such as synthetic data [34] or by pooling data from a larger numbers of sites. In addition, the expert labelers of the RSNA dataset annotated cases with post-operative collections as positive for ICH which accounts for the number of false positive cases with post-operative changes in our study. Adding a post-operative label to the training dataset would likely help reduce the number of false positives. ML model training and validation were performed on 5 mm slice thickness images and our clinical test dataset was composed of 2.5 and 5 mm slice thickness images. The model's performance may be further improved by incorporating prior imaging studies, taking into account the natural evolution of ICH, and refining model training continuously. The ML model was evaluated on historical data rather than on a prospective basis. Ideally, the ML model would be evaluated as part of a prospective controlled trial at multiple institutions with different CT scanners and imaging protocols. If incorporated as part of a triaging system, such a study could help evaluate the impact on report turn-around time and patient outcomes. Traditionally studies have evaluated model performance on the basis of a confusion matrix, accuracy, sensitivity, and specificity which fails to take into account the impact of incorrect classification on patient outcomes particularly since the impact of false negative and positive predictions can be quite asymmetric. We hope this study can help lay the foundation for future investigators to examine these issues.

*Data availability*

The publicly available RSNA Intracranial Hemorrhage CT dataset used for model training is available at https://www.kaggle.com/c/rsna-intracranial-hemorrhage-detection/data. The external validation dataset is not publicly available. Model output and ground truth labels are available upon reasonable request by contacting the corresponding author.

*Code availability*

The source code used in this project can be made available on reasonable request by contacting the corresponding author.

*Acknowledgments*

The authors would like to acknowledge the contributions of Blair Jones and Dr. Zsolt Zador.

*Author Contributions*

Guarantor: EC had full access to the study data and takes responsibility for the integrity of the complete work and the final decision to submit the manuscript. Study concept and design: HS, EC, MM. Acquisition, analysis, or interpretation of data: All. Drafting of the manuscript: HS, EC. Critical revision of the manuscript: All. Obtaining funding: N/A. Administrative or technical support: HS, EC, HML. Supervision: EC, MM

*Competing Interests*

The authors report no conflicts of interest.